\documentclass[twocolumn,10pt,conference]{IEEEtran}

\usepackage[T1]{fontenc}
\usepackage{txfonts}
\usepackage{graphicx}

\newtheorem{lemma}{Lemma}
\newtheorem{proposition}[lemma]{Proposition}
\newtheorem{theorem}[lemma]{Theorem}
\newtheorem{corollary}[lemma]{Corollary}
\newtheorem{remark}[lemma]{Remark}
\newtheorem{definition}[lemma]{Definition}

\IEEEoverridecommandlockouts
\begin{document}
\hypersetup{pdftitle={New Asymptotic Metrics for Relative Generalized Hamming Weight},pdfauthor={Ryutaroh Matsumoto},pdfkeywords={linear code, ramp secret sharing, generalized Hamming weight, Gilbert-Varshamov bound}}
\title{New Asymptotic Metrics for Relative Generalized Hamming Weight\thanks{%
This manuscript was accepted for presentation in 2014 IEEE
International Symposium on Information Theory, June 29--July 4, 2014 
Hawai'i Convention Center --- Honolulu, HI, USA.}}
\author{\IEEEauthorblockN{Ryutaroh Matsumoto\IEEEauthorrefmark{1}}
\IEEEauthorblockA{\IEEEauthorblockA{\IEEEauthorrefmark{1}Department of Communications and Computer Engineering, Tokyo Institute of Technology, 152-8550 Japan}}}
\date{\today}
\maketitle
\thispagestyle{plain}
\begin{abstract}
It was recently shown that RGHW (relative generalized
Hamming weight) exactly expresses the security of
linear ramp secret sharing scheme.
In this paper we determine the true value of the asymptotic
metric for RGHW previously proposed by Zhuang et al.\ in 2013.
Then we propose new asymptotic metrics useful for
investigating the optimal performance of linear ramp secret
sharing scheme constructed from a pair of linear codes.
We also determine the true values of the proposed metrics
in many cases.
\end{abstract}

\section{Introduction}
It was shown in \cite{chen07,cruz10} that
any pair of nested linear codes $C_2 \subset C_1 \subset \mathbf{F}_q^n$
can be used for constructing a linear ramp
secret sharing scheme \cite{blakley85,yamamoto86}.
Recently, Kurihara et al.\ \cite{kurihara12}
showed that the smallest number of shares required for
an adversary to illegitimately obtain at least $t \log_2 q$ bits
of information is exactly expressed by the $t$-th
relative generalized Hamming weight 
(RGHW) of $C_1^\perp \subset C_2^\perp$
proposed by Luo et al.\  \cite{luo05}.
In order to clarify how good secret sharing schemes can be
constructed from linear codes,
it is indispensable to discover existential bounds on RGHW
similar to the Gilbert-Varshamov bound.
Toward this direction, Zhuang et al.\ \cite{zhuang13} and
this author \cite{matsumoto13}
proposed such existential bounds.

In \cite{zhuang13}, Zhuang et al.\
proposed an asymptotic metric of RGHW
that is the highest information rate of a larger code $C_1$ of length $n$
with specified RGHW $n\delta$, and
derived lower bounds on their proposed metric.
Firstly we show that the true value of their metric is $1-\delta$
in Theorem \ref{th2}.
Then we argue that their asymptotic metric is a bit inconvenient
for investigating the asymptotically optimal
linear ramp secret sharing scheme mainly because 
the positive information rate of information leakage
is not taken into account.
To overcome the inconveniences we propose new asymptotic metrics
for RGHW as Definitions \ref{def1} and \ref{def2}
and determine their true values in many cases in
Corollary \ref{coro1} and Theorem \ref{th4}
by using \cite{matsumoto13}.

\section{Preliminaries}
For $C_2 \subset C_1 \subset \mathbf{F}_q^n$,
the $t$-th RGHW is defined as \cite{luo05}
\begin{equation}
M_t (C_1, C_2) = \min_{I \subseteq \{1,\ldots, n\}}\{ |I| \mid 
\dim C_1 \cap V_I - \dim C_2 \cap V_I \geq t \}, \label{eq31}
\end{equation}
where $V_I = \{ (x_1$, \ldots, $x_n) \in \mathbf{F}_q^n \mid
i \notin I \Rightarrow x_i = 0 \}$ for a subset $I$ of the
index set $\{1$, \ldots, $n\}$. The author proposed
the following bound:
\begin{proposition}\label{prop1}\cite[Corollary 3]{matsumoto13}
For integers $a,u,v,w$, define
\begin{eqnarray}
N_1 (w,u) &=& \frac{\prod_{i=0}^{u-1} q^w - q^i}{\prod_{i=0}^{u-1}q^u - q^i}
\label{eq:n1}\\
N_2(w,u,v) &=& \frac{\prod_{i=0}^{v-1} q^w - q^{u+i}}{\prod_{i=0}^{v-1}
q^v - q^i}\label{eq:n2}\\
N_3(w,u,v,a) &=& N_1(u,a) \cdot N_2(w-a,u-a, v-a). \label{eq:n3}
\end{eqnarray}
There exists a nested pair $C_2 \subset C_1 \subset \mathbf{F}_q^n$
of linear codes with $\dim C_1 = k_1$ and $\dim C_2=k_2$
such that $M_t(C_1,C_2) \geq d$ if\footnote{%
$t \leq k_1-k_2-1$ was forgotten in \cite[Corollary 3]{matsumoto13}.}
$1 \leq t \leq k_1-k_2-1$ and
\begin{eqnarray}
&& {n \choose d} \sum_{b=t+1}^{k_1-k_2} \sum_{a=0}^{\min\{d-b,k_1-b,k_2\}}
N_1(d,a) \cdot N_2(n-a, d-a, k_2-a) \nonumber\\
&&\cdot N_3(n-k_2, d-a, k_1-k_2, b)\nonumber\\
&<& N_1(n,k_2) \cdot N_1 (n-k_2, k_1-k_2). \label{eq:main}
\end{eqnarray}
\end{proposition}
\begin{IEEEproof}
See \cite{matsumoto13}.
\end{IEEEproof}

In order to derive asymptotic expressions, we need
the following lemmas.
\begin{lemma}\cite[Corollary 2]{helleseth95}
Define $\pi(q) = \prod_{i=1}^\infty (1-q^{-i})$. Then we have
\begin{equation}
\pi(q)q^{u(w-u)} \leq N_1 (w,u) \leq \pi(q)^{-1}q^{u(w-u)}.
\label{eq1}
\end{equation}
\end{lemma}
\begin{lemma}
\begin{eqnarray}
N_2 (w,u,v) &\leq& \pi(q)^{-1}q^{v(w-v)} \label{eq3}\\
N_3 (w,u,v,a) &\leq& \pi(q)^{-1} q^{u(u-a)}q^{(v-a)(w-v)}. \label{eq4}
\end{eqnarray}
\end{lemma}
\begin{IEEEproof}
Equation (\ref{eq3}) follows from (\ref{eq1}) and
$N_2(w,u,v) \leq N_2(w,0,v) = N_1(w,v)$.
Equation (\ref{eq4}) follows from (\ref{eq1}) and (\ref{eq3}).
\end{IEEEproof}

In order to evaluate the binomial coefficient asymptotically,
we use the following lemma.
\begin{lemma}\cite[Example 11.1.3]{cover06}
Define $H_q(x) = -x \log_q x - (1-x) \log_q (1-x)$.
Then
\begin{equation}
\frac{1}{n+1} q^{n H_q(m/n)} \leq {n \choose m} \leq q^{n H_q(m/n)}.
\label{eq2}
\end{equation}
\end{lemma}
We shall use the Singleton bound in proofs of
Lemma \ref{lem3}, Theorem \ref{th2} and Corollary \ref{coro1}.
\begin{proposition}[Singleton bound]\cite{zhuang13}
For any $C_2 \subset C_1 \subset \mathbf{F}_q^n$ we have
$M_t(C_1$, $C_2) + \dim C_1 \leq n + t$.
\end{proposition}

\section{Asymptotic Metrics for RGHW}
Zhuang et al.\ \cite{zhuang13} introduced
the following asymptotic metric
$\alpha_q^{(t)}(\delta) = \limsup_{n\rightarrow \infty}
\max \{ \dim C_1/n \mid$ there exists some $C_2 \subset C_1 \subset
\mathbf{F}_q^n$ such that
$M_t(C_1$, $C_2)/n \geq \delta \}$.
In order to investigate the best realizable
performance of secret sharing asymptotically, there are the following
inconveniences in $\alpha_q^{(t)}(\delta)$.
\begin{enumerate}
\item\label{l1} $t$ does not grow with $n$ in $\alpha_q^{(t)}(\delta)$
and $t/n$ tends to zero as $n\rightarrow \infty$.
In the context of secret sharing, $t/n$ represents the rate
of information leaked to an adversary who has collected
shares illegitimately.
In the ramp secret sharing scheme \cite{blakley85,yamamoto86},
it is important to analyze nonzero information leakage
as well as the zero information leakage.
The definition of $\alpha_q^{(t)}(\delta)$ cannot enable
such analysis.
\item\label{l2} The definition of $\alpha_q^{(t)}(\delta)$
does not take $\dim C_2$ into account.
The information rate of secret messages is $(\dim C_1 - \dim C_2)/n$
in secret sharing, but $\alpha_q^{(t)}(\delta)$ does not
enable analysis of information rate.
\item\label{l3} We shall show in Theorem \ref{th2}
$\alpha_q^{(t)}(\delta) = 1-\delta$, which makes
$\alpha_q^{(t)}(\delta)$ independent of $t$ in addition to $\dim C_2$.
\end{enumerate}
\begin{lemma}\label{lem3}
For any $0 \leq k_2 < k_1 \leq n$ and any prime power
$q$, if $t = k_1 -k_2$ then there exists $C_2 \subset C_1 \subset
\mathbf{F}_q^n$ such that $M_t(C_1$, $C_2) = n+t-k_1$.
\end{lemma}
\begin{IEEEproof}
Let $C_2$ be $V_{J}$ with $J = \{1$, $2$, \ldots, $k_2\}$
if $k_2 \geq 1$ and $C_2 = \{0\}$ if $k_2=0$.
Let $D$ be any $(k_1-k_2)$-dimensional space whose support is
$\{k_2+1$, \ldots, $n\}$. Clearly we have $D \cap C_2 = \{0\}$.
Let $C_1 = D + C_2$.

In order for $V_I$ in (\ref{eq31}) to satisfy
$\dim C_1 \cap V_I - \dim C_2 \cap V_I \geq t$,
$V_I$ must contain $D$, which makes $|I| \geq n-k_2 = n+t-k_1$.
This shows $M_t(C_1$, $C_2) \geq n+t-k_1$. On the other hand,
$M_t(C_1$, $C_2)$ cannot be larger than $n+t-k_1$ because of
the Singleton bound for RGHW \cite{zhuang13}.
\end{IEEEproof}

\begin{theorem}\label{th2}
\begin{equation}
\alpha_q^{(t)}(\delta) = 1-\delta.
\label{eq24}
\end{equation}
\end{theorem}
\begin{IEEEproof}
Let $d = \lfloor n\delta \rfloor$,
$k_1 = n+t - d$, $k_2 = n-d$.
By Lemma \ref{lem3} there exist
$C_2 \subset C_1 \subset \mathbf{F}_q$ such that
$\dim C_1 = n+t - \lfloor n\delta \rfloor$,
$\dim C_2 = n - \lfloor n\delta \rfloor$,
and $M_t(C_1$, $C_2) = \lfloor n\delta \rfloor$.
Letting $n\rightarrow \infty$ shows
$\alpha_q^{(t)}(\delta) \geq 1-\delta$. 
The Singleton bound shows
$\alpha_q^{(t)}(\delta) \leq 1-\delta$.
\end{IEEEproof}

In order to overcome these inconveniences,
we propose the following asymptotic metric for RGHW:
\begin{definition}\label{def1}
$\delta_{q} (\tau$, $R_1$, $R_2)= \limsup_{n\rightarrow \infty}
\max \{ d/n \mid$ there exists some $
C_2 \subset C_1 \subset \mathbf{F}_q^n$ such that $
t \leq n\tau$, $\dim C_1 \geq nR_1$,
$\dim C_2 \leq nR_2$, $M_t(C_1$, $C_2)\geq d \}$.
\end{definition}
The proposed new metric has the following operational
meanings:
$\delta_{q} (\tau, R_1, R_2)$ is the largest fraction
of shares in secret sharing constructed from
a code $C_1$ with information rate $R_1$ and $C_2$ with $R_2$
with which an adversary with arbitrary $n\delta_{q}$ shares
has at most $\tau \log_2 q$ bits of information per codeword
symbol.

\section{Asymptotic Analysis}
\begin{theorem}\label{th3}
Fix $0\leq R_1 \leq 1$, $0\leq \delta \leq 1$ and
$0 < \tau \leq \min\{R_1$, $\delta\}$.
If
\begin{eqnarray}
R_1 + \delta < 1 + \tau,
\label{eq13}
\end{eqnarray}
then for any $0 \leq R_2 \leq R_1-\tau$, any prime power
$q$, and sufficiently large $n$,
there exist
$C_2 \subset C_1 \subset \mathbf{F}_q^n$
such that $\dim C_1 = \lfloor nR_1 \rfloor$,
$\dim C_2 = \lceil nR_2 \rceil$,
and $M_{\lceil n\tau \rceil}(C_1$, $C_2) \geq \lfloor n \delta \rfloor$.
\end{theorem}
\begin{IEEEproof}
We shall show that (\ref{eq13}) implies (\ref{eq:main}) when
$n$ is large.
Fix $R_2$ and $q$ as assumed in the theorem.
Observe that
\begin{equation}
\sum f(j) \leq \max f(j) \times \textrm{the number of terms in }\sum.
\label{eq21}
\end{equation}
By using this idea and (\ref{eq1})--(\ref{eq2}), we see the left hand side of
(\ref{eq:main}) is less than or equal to
\begin{eqnarray}
&&\underbrace{(k_1-k_2-t)(1+\min\{d-b,k_1-b,k_2\})\pi(q)^{-4}}_{=\textrm{poly}(n)}
\label{eq22}\\
&&\times \textrm{exponential function of } n. \nonumber
\end{eqnarray}
We also see that the right hand side of (\ref{eq:main})
is also an exponential function of $n$.
In order to seek a sufficient condition for (\ref{eq:main}) when
$n$ is large, we can ignore $\textrm{poly}(n)$ in
(\ref{eq22}), take $\log_q$, and divide
it by $n^2$ (not $n$). Then by using (\ref{eq1})--(\ref{eq2}) and (\ref{eq21})
we see that
\begin{eqnarray}
&& \frac{k_2}{n}\left(1-\frac{k_2}{n}\right) +
\left(\frac{k_1}{n}-\frac{k_2}{n}\right)
\left(1-\frac{k_1}{n}\right)\nonumber\\
&>&
\max_{t/n +1/n\leq b/n \leq k_1/n-k_2/n, 0\leq a/n\leq \min\{ d/n-b/n, k_1/n-b/n,k_2/n\}} \nonumber\\
&&[
\frac{H_q(d/n)}{n} + a/n(d/n-a/n)
+(k_2/n-a/n)(1-k_2/n) \nonumber\\
&&+
(b/n)(d/n-a/n-b/n) + (k_1/n-k_2/n-b/n)(1-k_1/n)]\nonumber\\
&&
\label{eq23}
\end{eqnarray}
is a sufficient condition for (\ref{eq:main}) when $n$ is large.
Observe that the maximum is always achieved at $b=t+1$ and
we can substitute $b$ by $t+1$.
By identifying $R_1$, $R_2$, $\alpha$, $\delta$, $\tau$ with
$k_1/n$, $k_2/n$, $a/n$, $d/n$, $t/n$ we see that
\begin{eqnarray}
&& R_2(1-R_2) + (R_1-R_2)(1-R_1)\nonumber\\
&>&
\frac{H_q(\delta)}{n} + \max_{0\leq \alpha\leq \min\{ \delta-\tau, R_1-\tau,R_2\}} [
 \alpha(\delta-\alpha)
+(R_2-\alpha)(1-R_2) \nonumber\\
&&+
\tau(\delta-\alpha-\tau) + (R_1-R_2-\tau)(1-R_1)],
\label{eq41}
\end{eqnarray}
is a sufficient condition for (\ref{eq23}) when
$n$ is sufficiently large.
Since $\delta\geq \tau$,
we see that the maximum in (\ref{eq41}) is achieved at $\alpha=0$.
Substituting $\alpha=0$ into (\ref{eq41}) yields
\begin{eqnarray}
&& R_2(1-R_2) + (R_1-R_2)(1-R_1)\nonumber\\
&>&
\frac{H_q(\delta)}{n}+ R_2(1-R_2) +
\tau(\delta-\tau) + (R_1-R_2-\tau)(1-R_1).\nonumber\\
&&\label{eq42}
\end{eqnarray}
Subtracting $R_2(1-R_2) + (R_1-R_2)(1-R_1)$ from both sides
yields
\begin{equation}
0 > \frac{H_q(\delta)}{n}+\tau (R_1 + \delta - 1 - \tau). \label{eq43}
\end{equation}
Since we have assumed $\tau > 0$ we can divide (\ref{eq43}) by
$\tau$, ignore $\frac{H_q(\delta)}{n}$ and obtain (\ref{eq13}).
\end{IEEEproof}
\begin{remark}
One can deduce Theorem \ref{th3} not only from 
Proposition \ref{prop1} but also from
\cite[Theorem 2 or 3]{zhuang13}.
\end{remark}
\begin{corollary}\label{coro1}
If $\tau > 0$ or $\tau = R_1 - R_2$ then
\[
\delta_{q}(\tau, R_1, R_2) = 1+\tau - R_1.
\]
\end{corollary}
\begin{IEEEproof}
For $\tau > 0$, Theorem \ref{th3} shows $\delta_{q}(\tau, R_1, R_2) \geq 1+\tau - R_1$.
For $\tau = R_1 - R_2$, Lemma \ref{lem3} does it.
The opposite inequality follows from the Singleton bound for
RGHW.
\end{IEEEproof}

When $\tau=0$ and $R_1 > R_2$, the situation is unclear.
In order to investigate that case, we introduce another definition.
\begin{definition}\label{def2}
Let $t$ be a positive integer.
$\delta^0_{q} (t$, $R_1$, $R_2)= \limsup_{n\rightarrow \infty}
\max \{ d/n \mid$ there exists some $
C_2 \subset C_1 \subset \mathbf{F}_q^n$ such that 
$\dim C_1 \geq nR_1$,
$\dim C_2 \leq nR_2$, $M_t(C_1$, $C_2)\geq d \}$.
\end{definition}

\begin{theorem}\label{th4}
\begin{eqnarray}
\delta^0_{q} (t, R_1, R_1) &=& 1-R_1 \label{eq101}\\
\delta^0_{q} (t, R_1, R_2) &\geq & 1 - \frac{H_q(\delta)}{t}-R_1. \label{eq102}
\end{eqnarray}
\end{theorem}
\begin{IEEEproof}
Equation (\ref{eq101}) follows from Lemma \ref{lem3}.
To show (\ref{eq102}), by substituting $\tau=t/n$
 to (\ref{eq43})
and multiplying $n$ to both sides, we obtain
\[
0 > H_q(\delta)+ t (R_1 + \delta - 1 - t/n).
\]
Dividing it by $t$ yields
\[
\delta < 1 +\frac{t}{n} - \frac{H_q(\delta)}{t} - R_1
\]
as a sufficient condition for (\ref{eq:main}) when $n$ is large,
and we see (\ref{eq102}).
\end{IEEEproof}

\begin{figure}
\includegraphics[width=\linewidth]{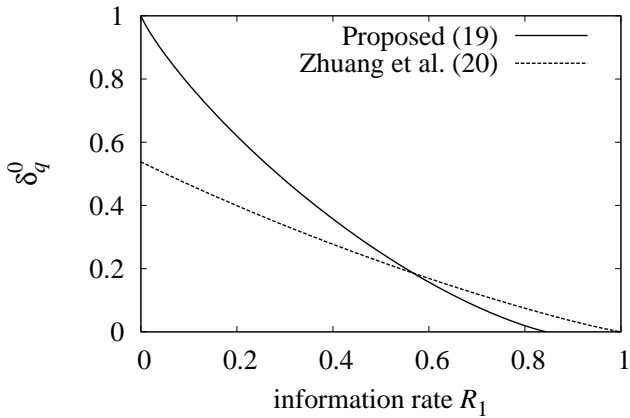}
\caption{Comparison of lower bounds on $\delta_q^0(t, R_1, R_2)$
for $q=4$ and $t=2$.}\label{fig1}
\end{figure}

As $t\rightarrow \infty$ (\ref{eq102}) tends to
$1-R_1$, which seems consistent with Corollary \ref{coro1}.
Zhuang et al.\ \cite[Eq.\ (25)]{zhuang13} showed
\begin{eqnarray}
\delta_q^0(t, R_1, R_2) &\geq& \max \{ \delta \mid
 R_1 \geq 1-\delta + \frac{\delta}{t}  \log_q \frac{\delta}{1-q^{-t}}\nonumber\\
&&+ \frac{1-\delta}{t}\log_q (1-\delta)
\}. \label{eq103}
\end{eqnarray}
As $t\rightarrow \infty$ (\ref{eq103}) also tends to
$1-R_1$. (\ref{eq102}) and (\ref{eq103}) are compared in
Fig.\ \ref{fig1} for $q=4$ and $t=2$.

\section{Conclusion}
In this paper we determined the true value of the asymptotic
metric for RGHW proposed by Zhuang et al.\ \cite{zhuang13}.
Then we proposed new asymptotic metrics useful for
analysis of \emph{ramp} secret sharing schemes. We also determined
the true values of the proposed metrics in many cases.

\section*{Acknowledgment}
The author would like to thank anonymous reviewers of IEEE ISIT 2014
providing comments helpful to improve this manuscript.
He would also like to thank Olav Geil, Diego Ruano, Stefano
Martin in Aalborg University, and
Yuta Hasegawa in Tokyo Institute of Technology
for stimulating discussions.
This research is partly supported by the JSPS Grant  
 Nos.\ 23246071 and 26289116.


\end{document}